\documentclass[final,3p,times,twocolumn]{elsarticle}

\usepackage{amssymb,amsmath}
\usepackage{bm}

\usepackage{booktabs}
\usepackage{braket}
\usepackage{url}
\usepackage[english]{babel}
\usepackage{graphics}
\usepackage{tabularx}
\usepackage{rotating}
\usepackage{float}

\graphicspath{{./figures/}}

\journal{Physics Letters A}

\begin{document}

\begin{frontmatter}

\title{Few-body quantum method in a $d$-dimensional space}


\author[1]{E. Garrido}
\ead{e.garrido@csic.es}
\author[2]{A.S. Jensen}
\author[3]{R. \'{A}lvarez-Rodr\'{\i}guez}

\address[1]{Instituto de Estructura de la Materia, IEM-CSIC, Serrano 123, E-28006 Madrid, Spain}
\address[2]{Department of Physics and Astronomy, Aarhus University, DK-8000 Aarhus C, Denmark} 
\address[3]{Escuela T\'{e}cnica Superior de Arquitectura, Universidad Polit\'{e}cnica de Madrid,
Avda. Juan Herrera 4, 28040 Madrid, Spain}

\begin{abstract}
In this work we investigate the continuous confinement of quantum systems from three to
two dimensions. Two different methods will be used and related. In the first one the
confinement is achieved by putting the system under the effect of an external field.
This method is conceptually simple, although, due to the presence of the external field,
its numerical implementation can become rather cumbersome, especially when the system is
highly confined. In the second method the external field is not used, and it simply considers
the spatial dimension $d$ as a parameter that changes continuously between the ordinary integer values.
In this way the numerical effort is absorbed in a modified strength of the centrifugal barrier.
Then the technique required to obtain the wave function of the confined system is precisely the same
as needed in ordinary three dimensional calculations without any confinement potential.
The case of a two-body system squeezed from three to two dimensions is considered, and used
to provide a translation between all the quantities in the two methods.
Finally we point out perspectives for applications on
more particles, different spatial dimensions, and other confinement potentials. 
\end{abstract}

\end{frontmatter}

\section{Introduction}

The spatial dimension crucially determines the quantum mechanical
properties of physical systems. In three dimensions a  finite attraction is 
necessary to provide a bound state, whereas in two dimensions any infinitesimal 
attraction is enough to support a bound state \cite{sim76,lan77,nie01,pet01}.  
This is most clearly seen for a system with one
spatial coordinate, as for one particle in a potential or for the
relative motion of two particles.  The centrifugal barrier for such a
simple system is zero or negative for $s$-waves in three and two
dimensions, respectively.

It is conceptually easy to envisage confinement to lower dimensions
than three, that is on planes or curves.  Also deformed microscopic
systems have been abundantly investigated, perhaps mostly in the
many-body nuclear problem \cite{zin14,sor02,tho08,hov18}, where symmetries can reduce
the treatment to fewer than three spatial dimensions.  Extreme
deformations simulate lower dimensions like pancakes or cigar shapes.
In nuclei these deformations are never extreme because they arise as
self-supported isolated structures without coupling to environment or 
other systems.

This is in sharp contrast to cold atoms and molecules where external
fields are used to confine systems into deformed and even all the way
to extreme lower dimensions \cite{kla17,nai17,lev14,san16a,san16b,chr18}.  The rapid
technical developments in conjunction with the qualitative change of
properties have prompted a number of both theoretical and experimental
investigations in one and two dimensions \cite{nie97,ped16,bar18}.
However, in practical cold gas experiments the extreme limits of integer  
dimensions are only approximately achieved. The techniques apply to strong
external fields which effectively can correspond to extreme deformations 
that only allow spatial extension in less than three dimensions.

Earlier attempts to study dimensional dependencies are available in
particle and mathematical physics \cite{lei75, gon92,dun99}, and in the
so-called $1/d$-expansion \cite{dor86,her92,dun93}. In these methods
the dimension $d$ is considered as an integer parameter, in such a way that for
some particular values of $d$ the quantum mechanical solutions are
available or readily obtained. Expanding the solutions around these
particular $d$-values it is then possible to extrapolate the solution
for the dimension of interest. The aim is to find correct converged
solutions through this expansion. In this case all those solutions for
$d$-values different from the one of interest have no physical meaning.

In this work the philosophy when introducing the dimension as a parameter is 
different. For practical reasons we focus here on dimensions between $2$ and $3$,
where solutions can be obtained directly and accurately. Non-integer values of $d$ 
have now a physical interpretation, namely, they describe the confinement of the system, 
which can evolve continuously from moving in a three-dimensional to a two-dimensional 
space. This formulation has been available for a number of years for isolated systems,
and especially detailed, but not 
exclusively, for two and three particles \cite{nie01}. This formulation is precisely as
efficient as ordinary integer dimension calculations, that is, as optimal as
possible.  However, the formalism is not connected to the available
practical possibilities of laboratory tests, which on the other hand
is conceptually straightforward by applying an external field to
squeeze by gradually imposing confining walls. 

The purpose of the present work is to provide the initial steps for a
general and practically useful translation between the efficient
non-integer $d$-formulation and the brute force method using an
external field.  For this we consider the relative motion of a
two-body system, which has turned out to be interesting in itself,
although first investigated as preparation for the much more
complicated three-body system.  Thus we first sketch the key points of
the two methods, and second we provide interpretation and unique and
universal translation between the methods. Finally we discuss the
perspectives of applications to more complicated systems.

\section{Formulation}
\label{sec2}

The hamiltonian for two interacting particles in an external field in
three spatial dimensions can be written as
\begin{eqnarray}
  H &=& T + V(r)  + V_{ext}(z,Z) \; , \label{eq1} \\
  T &=& T_r + T_R \; , \\ \label{eq3}
  V_{ext}(z,Z)  &=&  \frac{1}{2} \omega_z^2 (M Z^2 + \mu z^2)  \; ,
\end{eqnarray}
where we assume an external harmonic oscillator field in
the $z$-direction, $V_{ext}(z,Z)$, of frequency $\omega_z$, acting on
both particles and depending on the $Z$ and $z$ components of the 
center-of-mass and relative coordinates, $r$ and $R$, respectively, and on the total and 
reduced masses $M$ and $\mu$. We emphasize that the frequency is the same for both
particles to ensure separation in relative and center-of-mass
coordinates of both the external field and kinetic energy operator
$T$. The particle-particle interaction, $V(r)$, will be assumed to be short-range.

The center-of-mass motion can now be solved completely independently,
and in the following we shall concentrate on the relative coordinates.
It is usual to give the oscillator frequency $\omega_z$ in terms of the
oscillator length $b_{ext}=\sqrt{\hbar/(\mu\omega_z)}$, in such a way that the
external potential can be written as 
\begin{equation}
V_{ext}(z)=\frac{\hbar^2}{2\mu}\frac{z^2}{b_{ext}^4},
\label{eq4}
\end{equation}
with $z=r\cos\theta$, where $\theta$ is the usual polar angle.

Note that the external potential (\ref{eq4}) has a non-central character, which means that,
as soon as $V_{ext}\neq 0$ the relative orbital angular momentum is not
a good quantum number, and the total confined two-body wave function,
$\Psi_{b_{ext}}$, has to be obtained including a sufficient number of partial waves
such that convergence is achieved, i.e.,
\begin{equation}
\Psi_{b_{ext}}(\bm{r})=\sum_{\ell m} \frac{u_\ell(r)}{r} Y_{\ell m}(\theta,\varphi).
\label{wf}
\end{equation}
In particular, the smaller the oscillator length $b_{ext}$, the larger the effect of $V_{ext}$, and the higher 
the number of partial waves required. For $b_{ext}=\infty$, or $V_{ext}=0$, due to the central character
of the two-body interaction $V(r)$, the quantum number $\ell$ is conserved and the expansion above reduces
to one $\ell$-term.

The radial wave functions,
$u_\ell(r)$, are obtained as the solution of the coupled set of differential equations:
\begin{eqnarray}
\lefteqn{ \hspace*{-0.5cm}
\left[ \frac{\partial^2}{\partial r^2}-\frac{\ell(\ell+1)}{r^2}-
\frac{2\mu}{\hbar^2} (V(r)-E_{tot})
\right] u_\ell} \nonumber \\ && \hspace*{5mm}
-\frac{r^2}{b_{ext}^4}\sum_{\ell' m'}\langle Y_{\ell m}|\cos^2 \theta|Y_{\ell' m'}\rangle_\Omega
                            u_{\ell'}=0,
\label{sqeq}
\end{eqnarray}
where $E_{tot}$ is the total energy, and $\langle \rangle_\Omega$ indicates integration over the polar and azimuthal angles only.
Note as well that  $\langle Y_{\ell m}|\cos^2 \theta|Y_{\ell' m'}\rangle_\Omega \propto \delta_{m m^\prime}$, 
which implies that the orbital angular momentum projection $m$ along the $z$-axis remains as a good 
quantum number. 

In this work we shall consider relative $s$-waves between the particles when $V_{ext}=0$. This means that
for any value of $b_{ext}$, the quantum number $m$ will be equal to zero.
As a consequence, the dependence of the wave function (\ref{wf}) on the azimuthal angle $\varphi$ disappears.

This method is straightforward and simple, but generalization to three
and perhaps more particles is quickly much more time consuming.  To
circumvent this complication we turn to another formulation
\cite{nie01,ros17,ros19} where the external field is omitted altogether,
whereas the spherical two-body potential is maintained in $d$ dimensions.  The
relative Schr\"{o}dinger equation is then
\begin{equation}  
\label{schd}
\bigg[\frac{\partial^2}{\partial r^2} - \frac{(d-3)(d-1)}{4r^2} 
   - \frac{2\mu}{\hbar^2} (V(r) -E)\bigg] R_d(r) = 0,
\end{equation}
where a relative $s$-wave has been assumed, $E$ is the energy, and $R_d(r)$ is 
the reduced radial wave function.  The
idea is now to vary $d$ continuously between $d=3$ and $d=2$, where
the equation of motion in these two limits \cite{nie01} are correct as seen by
inspection of Eq.(\ref{schd}). This computation is much faster than by
using the external field, especially for small values of $b_{ext}$, where
Eq.(\ref{sqeq}) involves a large number of coupled differential equations. 

The general form of Eq.(\ref{schd}) for integer values of $d$ contains
an additional repulsive centrifugal term, i.e. $l_d(l_d+d-2)/r^2$, where $l_d$ 
is an integer, the angular
momentum quantum number for $d=3$ and its projection for $d=2$. 
The corresponding angular wave functions are generalized spherical harmonics or
related to Jacobi functions, where the coordinates are angles
\cite{nie01}. For $s$-waves such an angular dependence disappears.
However, we can imagine excited and non-spherical
states in both two and three dimensions.  This extension to partial
waves other than $s$-waves is definitely not trivial, since it
necessarily at least implies an interpretation of $d-1$ angular
variables, where $d-1$ may be a non-integer number.  Such a
generalization of the method is far beyond the scope of this report,
but it is not entirely inconceivable for example by maintaining the
azimuthal angle for $d{\in}[2,3]$, while gradually limiting the polar
angle to a smaller interval than from $0$ to $\pi$.

However, this research direction is not tempting, since the universal
behavior of quantum systems is closely related to $s$-waves (or
marginally also $p$-waves).  Furthermore, large scattering lengths are
necessary ingredients to have systems insensitive to the details of
the potentials.  Otherwise bound state wave functions would be pulled
into the interior of the attractive potential and therefore loosing
the universal character by being strongly dependent on the shape of
the potential.  For this reason, if a universal relation between the
squeezing parameter $b_{ext}$ and the dimension $d$ exists, the
restriction in Eq.(\ref{schd}) to relative $s$-waves appears very appropriate.

\begin{table}
\caption{Strengths, $S_g$ and $S_m$, for the three Gaussian (upper part of the table) and 
Morse (lower part of the table) potentials giving rise to the
three-dimensional scattering lengths $a_{3D}$. For each case we also give the
$s$-wave two-body binding energies in 2D and 3D, $E_{2D}$ and $E_{3D}$, and
the corresponding root-mean square radii $r_{2D}$ and $r_{3D}$.
All the energies are given in units of $\hbar^2/\mu b^2$ and the lengths in units of $b$, where $b$
is the range of either the Gaussian or the Morse potential.}
\label{tab1}
\begin{center}
\begin{tabular}{l|lll}
&Pot. I&Pot.II&Pot.III\\ \hline
$a_{3D}$          &2.033   &18.122&40.608\\ \hline
$S_g$     &$-2.71$ & $-1.43$ &$-1.38$ \\
$E_{2D}$ &$-0.908$&$-0.269$ &$-0.249$ \\
$E_{3D}$ &$-0.269$&$-1.651\cdot10^{-3}$&$-3.144\cdot10^{-4}$\\
$r_{2D}$ &0.926 &1.398&1.439 \\
$r_{3D}$ &1.508 &12.823&28.710\\ \hline
$S_m$     &$1.294$ & $0.474$ &$0.434$ \\
$E_{2D}$ &$-0.450$&$-7.394\cdot10^{-2}$ &$-6.088\cdot10^{-2}$ \\
$E_{3D}$ &$-0.189$&$-1.875\cdot10^{-3}$&$-3.325\cdot10^{-4}$\\
$r_{2D}$ &1.458 &2.739&2.947 \\
$r_{3D}$ &2.235 &12.870&28.741\\ \hline
\end{tabular}
\end{center}
\end{table}

\begin{figure}[t]
  \centering
\includegraphics[width=0.95\linewidth]{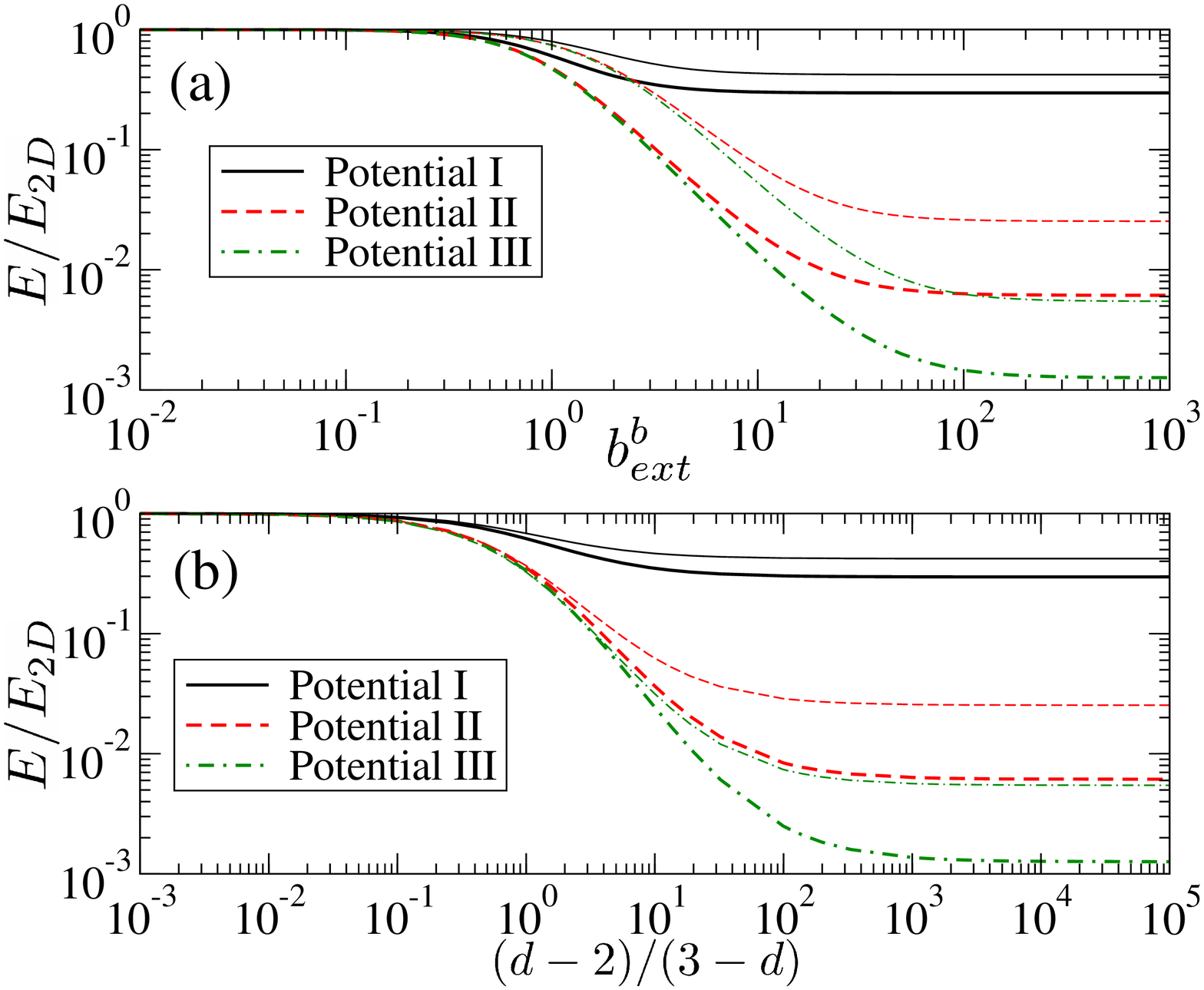}  
  \caption{Two-body energies, normalized to the $2D$-limit, for the Gaussian (thick curves)
  and Morse (thin curves) potentials in Table~\ref{tab1}. The $x$-axis for panels (a) and (b) are the
  parameters used in the two methods, that is $b_{ext}^b = b_{ext}/b$ and
  $(d-2)/(3-d)$, respectively.}
\label{fig1}     
\end{figure}

\section{Confinement length versus $d$}

The simplest characterizing quantities of any system are energy and
radius, which can be used to give the connection between the
parameters of the two methods, i.e., between $b_{ext}$ and $d$. 

Two different shapes for the two-body potential $V(r)$ will be used:
A Gaussian potential, $V(r)=S_ge^{-r^2/b^2}$, and a Morse-like potential, 
$V(r)=S_m(e^{-2r/b}-2e^{-r/b})$. For each of them we have chosen three 
cases, each supporting one weakly bound two-body state, and whose three-dimensional
(3D) identical scattering lengths, $a_{3D}$, are given in Table~\ref{tab1}. The strengths of the
Gaussian and Morse potentials, $S_g$ and $S_m$, corresponding to each of the
three $a_{3D}$ values, are also given in the table, together with the 2D and
3D binding energies, $E_{2D}$ and $E_{3D}$, and root-mean-square
radii, $r_{2D}$ and $r_{3D}$ for each case.
We simplify by measuring lengths in units of the potential range,
$b$, and energies in units of $\hbar^2/\mu b^2$.  

Note that the value of $a_{3D}$ increases significantly from
potential I to III, implying a decrease of the binding energy
$\left|E_{3D}\right|$. The strength of the Gaussian  and Morse potentials in the natural units in
Table~\ref{tab1} approaches the limit $S_g=-1.34$ and $S_m=0.405$, respectively,
corresponding to an infinitely large 3D scattering length. 
In this limit, the 2D scattering length, $a_{2D}^\infty$, for the Gaussian
and Morse potentials, takes the value 1.99 and 4.15, respectively, 
each in its corresponding length unit.

\begin{figure}[t]
\centering
\includegraphics[width=0.95\linewidth]{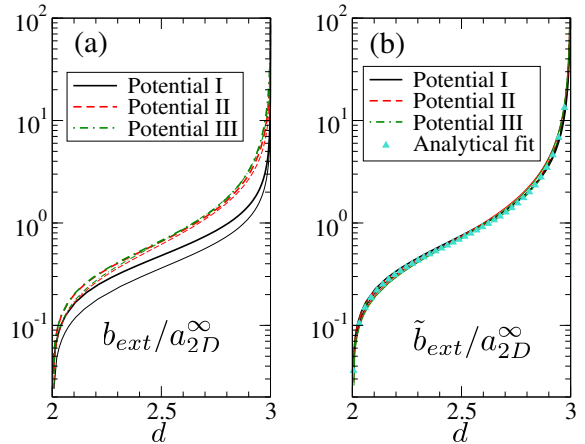}
\caption{The derived dependence of $b_{ext}/a_{2D}^\infty$ as function of $d$ in
  panel (a) for the potentials in Table~\ref{tab1}. The thick and thin curves are the results
  for the Gaussian and Morse potentials, respectively. The universal
  curve in panel (b), parameterized in Eq.(\ref{enerfit}), arises after
  use of the correction factor given  in Eq.(\ref{corr}). }
\label{fig2}     
\end{figure}

For the different potentials we then calculate the corresponding
energies with the two methods as functions, respectively of the
squeezing length $b_{ext}^b=b_{ext}/b$ and the $d$-parameter. The
corresponding results for the Gaussian (thick curves) and Morse (thin
curves) potentials are shown in Fig.~\ref{fig1} normalized to the
2D energy. For the calculation using the external harmonic oscillator
potential the energy $E$ is obtained after subtracting the zero point
energy in the squeezed $z$-direction from the
total computed energy in Eq.(\ref{sqeq}), i.e., $E=E_{tot}-\hbar \omega_z/2$.  By construction, all the curves reach unity in
the $2D$ limit, while the energies are much lower in $3D$, where the
values for both methods are $\hbar^2/(2\mu a_{3D}^2)$ for large
scattering lengths. 

The idea is now to relate $b_{ext}$ and the $d$-parameter by requiring
that the energies are identical. For this to be useful we must find a
model-independent translation.  As we have learned from nuclear halo physics
\cite{nie01,jen04,fre12}, and extensively as well in cold atomic gases
\cite{nai17,bra06,fre11}, the appearance of a universal, model-independent, behavior
of quantum systems is associated to the presence of relative $s$-waves and large scattering lengths.
This has been established as a universal parameter describing
properties of weakly bound states without reference to the responsible
short-range attraction. This is equivalent to the effective range
expansion of the low-energy scattering phase shift, where the leading
term is given by the scattering length.  The next order entering with
increasing energy depends on the range of the potential.  The same
properties are described by these quantities for disparate potentials.
For this reason, the translation between $b_{ext}$ and $d$ obtained from 
Fig.~\ref{fig1} for the potentials with large scattering length should be
very close to the desired universal relation between the two parameters. In fact,
the precise relation should be provided by a potential with infinitely large 
scattering length.

The results are shown in Fig.~\ref{fig2}a, where we show $b_{ext}/a^\infty_{2D}$ 
as function of $d$ for the three potentials for the Gaussian
(thick curves) and Morse (thin curves) cases, where $a_{2D}^\infty$
is the 2D scattering length obtained with a potential strength providing 
an infinite scattering length in 3D. This amounts to using the same length
unit for the two potential shapes, since, no matter the shape, the 
potential providing an infinite scattering length for $d=3$, gives rise
to a scattering length, in the unit $a_{2D}^\infty$, equal to 1 for $d=2$. 
As we can see, when the scattering length is large (potentials II and III), as expected, 
the short-range potential only has marginal influence,
since the wave function is almost universal and outside the
potential. The resulting translation is then already almost unique between
$b_{ext}/a^\infty_{2D}$ and $d$.

However, the uniqueness disappears when the
scattering length in three dimensions is comparable to the potential
range (potential I). This deficiency is to a large extent repaired by a correction factor, 
$\left(1+b_{ext}/a_{3D}\right)$, which however, for small $a_{3D}$ (potential I), still leaves small 
deviations for $d$ close to 2 and relatively small values of $b_{ext}$ comparable 
to the size of the system. The
final fine-tuning is achieved after the transformation:
\begin{equation}
\tilde{b}_{ext}=b_{ext} \left(1+\sqrt{\frac{b_{ext}^2+r_{2D}^2}{a_{3D}^2}} \right),
\label{corr}
\end{equation}
where $r_{2D}$ (Table~\ref{tab1}) is the root-mean-square radius for
the bound state in two dimensions. This form is found by trial and error and not by an analytic
derivation.  However, the idea is as for effective range expansion in
scattering theory, where the leading (universal) term is contained in
the scattering length, while the next to leading order term is related
to the range of the potential, which still is a universal average
property.

The resulting efficient modification results in Fig.~\ref{fig2}b with
the impressive reproduction for both potential shapes, as well as for small and large scattering
lengths.  The translation between $b_{ext}$ and $d$ only depends on
gross properties as the scattering lengths in two and three dimensions
and the mean-square-radius.  These three quantities are all available
by use of only calculations within the $d$-method. The translation is
universal, or model independent, in the sense that it is independent
of detailed properties of the potential.  This is demonstrated by the
results in Fig.~\ref{fig2}b for rather different short-range
potentials.

In Fig.~\ref{fig2}b we also show an analytical fit (triangles) that
reproduces very well the universal curve.  Since the curve is model
independent, the special form of the fitting function is unimportant
provided it is a sufficiently accurate fit which gives the connection
between $\tilde{b}_{ext}$ and $d$.  We have different options but one
possibility is
\begin{equation}  
 \frac{\tilde{b}_{ext}}{a_{2D}^\infty} = c_1 \frac{d-2}{3-d} + c_2 \tan\left((d-2)^{c_3}  \frac{\pi}{2} \right),
\label{enerfit}
\end{equation}
which is a combination of two functions, each of them being equal to zero at $d=2$, and to $\infty$
at $d=3$, and whose relative weight is used to fit the
curves between these two limits.  The computed fitting constants $c_k$ are $(c_1,c_2,c_3)=
(-0.207,0.535,0.592)$.

It seems perhaps superfluous to use two terms with the same asymptotic
behavior in both limits.  However, each of them cannot reproduce
intermediate dimensions. Analytic guidance is very difficult to find,
since it involves details about how the excited modes in the direction
perpendicular to squeezing are depopulated. Thus, not even the lowest
correction term to the $d=2$ limit is simple.  We leave this for
possible future investigations.

It is important to note that after doing the calculations using the root-mean-square
radii instead of the energy (Fig.~\ref{fig1}), the same results as shown 
in Fig.~\ref{fig2}, and therefore also in Eq.(\ref{enerfit}), are obtained. This is an indication 
of the robustness of the universal curve shown in Fig.~\ref{fig2}b.

\section{Wave function interpretation}

The calculation of the two-body wave function in $d$ dimensions is 
formulated in spherical coordinates, where, since $s$-waves are assumed,
the angles are absent. In particular, the corresponding wave function  $\Psi_d$ 
takes the form:
\begin{equation}
\Psi_d(r)=\frac{R_d(r)}{r^{(d-1)/2}}Y_d, \mbox{ with } 
Y_d=\left[\frac{\Gamma(\frac{d}{2}) }{2\pi^{d/2}} \right]^{1/2},
\label{eq8}
\end{equation}
where $R_d(r)$ is obtained from Eq.(\ref{schd}).  The root-mean-square
radius in $d$ dimensions is given by
\begin{equation}
r_d^2=\int_{0}^{\infty} r^2 \left|R_d(r)\right|^2 dr.
\label{eq9}
\end{equation}
This is a measure of the size of the wave function in $d$ dimensions,
which unfortunately has no immediate physical interpretation.  The
mathematically well-defined concepts in $d$ dimensions are not
directly applicable for computation of observables, which necessarily are
measured in 3D .  We therefore need an interpretation, or rather a 
translation, to be able to discuss ordinary three dimensional quantities.

In order to do so, we shall exploit the fact that we already have the
comparison to the ordinary calculations with an external field, and
the connection between $d$ and the external field parameter
(Fig.~\ref{fig2}).  It is clear that the squeezing performed in one
direction necessarily leads to a deformed structure.  It looks then
natural to try to reinterpret the total spherical wave function in $d$
dimensions, $\Psi_d$, as corresponding to a deformed three-dimensional
system.

To simulate the deformation produced by the external field, the most
appealing procedure is to deform by scaling the radial coordinate,
$r$, in the squeezed direction. We then take the usual Cartesian coordinates, 
$\{x,y,z\}$, in ordinary three dimensional space, and maintain the wave function,
$\Psi_d$, obtained in the $d$-calculation, but where its radial argument is
now substituted by
\begin{equation}
r\rightarrow \tilde{r} \equiv \sqrt{x^2+y^2+(z/s)^2} \equiv \sqrt{r_\perp^2+(z/s)^2} \;,
\label{scl}
\end{equation}
where $s$ is a scaling parameter which depends on the geometry of the
external potential, and in general might depend on $z$.  The
corresponding wave function, $\tilde{\Psi}_d \propto \Psi_d$, must be
normalized in the new three dimensional space, i.e.,
\begin{equation}
2\pi\int r_\perp dr_\perp dz |\tilde{\Psi}_d(r_{\perp},z,s)|^2=1 \;.
\label{eq11}
\end{equation}
It is clear that the limits of $d=2$ and $d=3$ are correct.  For instance, for $s=0$ only
$z=0$ is possible (otherwise $\tilde{r}=\infty$ and $\Psi_d(\infty)=0$) and we
are in the $d=2$ situation. Similarly, for $s=1$ the relative
coordinate $\tilde{r}$ is the usual radius in spherical coordinates, and we are
in the $d=3$ case.

The suitability of this procedure can be tested analytically by the
rather efficient one-gaussian approximation \cite{gaus1}.  We compare
the best one-gaussian wave functions for the same energy from the
external field, $R_{ext} \propto \exp(-r_{\perp}^2/(2b^2) -
z^2/(2b_z^2))$, and the $d$-calculations, $R_d(r) \propto
\exp(-r^2/(2b_d^2))$, where $R_{ext}$ is the wave function without
phase factor, and $r$ is scaled as in Eq.(\ref{scl}).  The maximum
overlap of unity is obtained for $b=b_d$ and $s= b_z/b_d$.  This
observation is very encouraging, although not surprising due to the
analogous one-gaussian assumptions, which often is rather efficient
and therefore making it even more appealing to exploit.

\begin{figure}[t]
\centering
\includegraphics[width=0.95\linewidth]{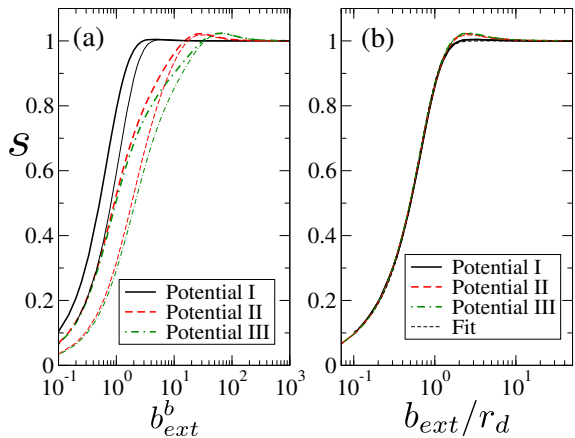}
\caption{The scaling parameter $s$ as a function of $b_{ext}^b=b_{ext}/b$ (panel a) and as a function
of $b_{ext}/r_d$ (panel b) for the potentials in Table~\ref{tab1}. The thick and thin curves
correspond to the Gaussian and Morse potentials, respectively. In panel (b) the
curve denoted as ``Fit'' (thin-dashed) corresponds to the analytical fit in Eq.(\ref{eq12}). }
\label{fig4}     
\end{figure}

\section{Wave function parameterization}

We generalize by comparing the wave functions from the $d$-calculation
and the full external field, $\Psi_{b_{ext}}(\bm{r})$ in Eq.(\ref{wf}), where
$r^2=r_\perp^2+z^2$.  The overlap between these functions is defined by 
\begin{equation}
{\cal O}(d,s)=2\pi \int r_\perp dr_\perp dz \tilde{\Psi}_d(r_\perp,z,s) \Psi_{b_{ext}}(\bm{r}) \; ,
\label{eq10}
\end{equation} 
where $s$ is adjusted to produce maximum overlap for the given $d$,
which in turn is universally related to $b_{ext}$ through Eq.(\ref{enerfit}),
see Fig.~\ref{fig2}b for accuracy.

The computed values of the scale parameter $s$ for the potentials used
in this work are shown in Fig.~\ref{fig4}a as a function of
$b_{ext}^b$. The larger the squeezing the smaller the scale parameter
which, as anticipated, varies between $0$ and $1$.  Only in the region
of little squeezing $s$ takes values slightly larger than 1. This fact
indicates that in this region ($d$ very close to 3) the interpretation
of the $d$-wave function as $\tilde{\Psi}_d(r_\perp,z,s=1)$ gives rise
to a state with the particles a bit too confined along the $z$-axis,
in such a way that maximization of the overlap (\ref{eq10}) requires a
small release of the confinement by means of a scale factor bigger
than 1.

The differences between the curves shown in Fig.~\ref{fig4}a are
related to the size of the two-body system.  For both, the Gaussian
and Morse shapes, potential III is the one
with the largest scattering length in 3D, and therefore the one
describing the largest two-body bound state. For this reason, the
curve corresponding to this potential is the first one feeling the
squeezing, i.e., it is the first one for which $s$ deviates from 1
when the squeezing parameter $b_{ext}^b$ is made smaller and
smaller. For the same reason the second potential feeling the
squeezing is potential II, and for potential I the deviation from
$s=1$ starts for even smaller values of $b_{ext}^b$. A simple way to
account for these size effects is to plot the scale parameter $s$ as a
function of $b_{ext}/r_d$, where $r_d$ is the rms radius of the
system for dimension $d$ as given in Eq.(\ref{eq9}).

The result is shown in Fig.~\ref{fig4}b, and, as we can see, all the
curves collapse into a single universal curve, which therefore permits
to relate the scale parameter $s$ and the squeezing parameter
$b_{ext}$ (or $d$ through Eq.(\ref{enerfit})) independently of the
potential. Only some small discrepancy is observed for potential I in
the region of little squeezing, where $s>1$.  The emerging universal
curve is very satisfying. We emphasize the crucial point that the
universal curve is defined and obtained entirely within the practical
$d$-calculation.

A simple analytic fit to the universal curve in Fig.~\ref{fig4}b is given by
\begin{equation}
\frac{1}{s(x)}=1+c_1\frac{e^{-c_2 x}}{x^{c_3}}
; \hspace*{1cm} x=\frac{b_{ext}}{r_d},
\label{eq12}
\end{equation}
with $c_1=2.75$, $c_2=2.92$, and $c_3=0.64$, which gives rise 
to the fitted and hardly distinguishable thin-dashed curve in Fig.~\ref{fig4}b.
The transition between $d=2$ and $d=3$ is exponentially fast in the region where 
the squeezing length is comparable to the size of the system.
The result in Eq.(\ref{eq12}) is again, as Eq.(\ref{enerfit}), an accurate but somewhat
arbitrary fitting function with precisely the same difficulties in
deriving analytically.

It is important to recall the parameter $s$ has been obtained for each
$d$ such that the overlap in Eq.(\ref{eq10}) is maximum. Obviously,
for $d=3$ the two wave functions involved in Eq.(\ref{eq10}) are
solutions of the same equation, and therefore the overlap is maximum 
and exactly 1 for $s=1$. However, under a large squeezing situation
the equations giving rise to $\tilde{\Psi}_d$ and $\Psi_{b_{ext}}$ 
are superficially very different, and therefore the overlap, even if maximized, 
might differ from 1.

\begin{figure}[t]
\centering
\includegraphics[width=0.95\linewidth]{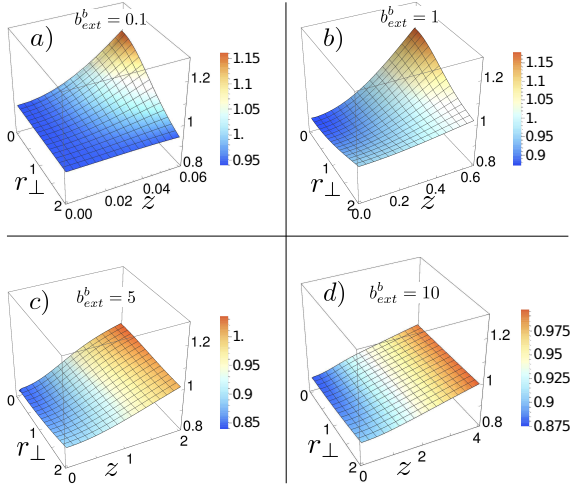}
\caption{For the Gaussian potential II, ratio between the wave functions $\Psi_{b_{ext}^b}(r)$ and $\tilde{\Psi}_d(r_\perp,z,s)$
for four different values of $b_{ext}^b$, as a function of $r_\perp$ and $z$ in units of $b$. The maximum value shown for the $z$-coordinate is
given by  $\langle z^2 \rangle^{1/2}$.}
\label{fig5}     
\end{figure}

Also, for $d=2$, both wave functions describe a spherical two-dimensional
structure obtained with the same interaction, and can
correspondingly be expected identical.  This is confirmed when $s$ is
chosen to maximize the overlap, which even in the most unfavorable
computed case ($b_{ext}^b=0.1$, i.e., $d$ and $s$ approaching 2 and 0,
respectively) is numerically found equal to 0.993, 0.984, and
0.984, for the Gaussian potentials I, II, and III, and 0.989, 0.980, and
0.977 for the Morse potentials I, II, and III, respectively.  This result
supports the interpretation of $\Psi_d$ described above for the
intermediate interval of $d$ between $2$ and $3$.

Instead of the very stable overlap value, we show in Fig.~\ref{fig5},
for the Gaussian potential II, the more sensitive ratio 
$\Psi_{b_{ext}}/\tilde{\Psi}_d$ as a function of $r_\perp$ and $z$ (in units of $b_b$). 
Panels $a$, $b$, $c$, and $d$ show,
respectively, the ratio obtained for squeezing parameters covering the
whole range of dimensions under investigation, that is
$b_{ext}^b=0.1$, $b_{ext}^b=1.0$, $b_{ext}^b=5.0$, and
$b_{ext}^b=10.0$. For each of the cases, the plots extend to the largest $z$-value taken
to be $\langle z^2\rangle^{1/2}$, where the wave functions only below this value differ
significantly from zero. The scale in the four panels has been chosen to emphasize the
differences between them. In any case, the ratio in Fig.~\ref{fig5} is rather stable and
close to $1$, until the far corner where each wave function is vanishingly small.
For this reason, these deviations are of no consequence, since they
are obviously inaccurate as obtained in the limit of zero divided by zero.
The same level of similarity (not shown) is found for the other
potentials. This is a conclusive proof of our interpretations.

\section{Procedure in $d$ dimensions}

The interpretation of $\Psi_d$ allows full calculation of any desired
observable, entirely within the efficient $d$-calculation.  In other
words the otherwise undefined angular part is now well-defined through
the translation and the universal curves.  For instance, we can
compute the partial wave content in the usual way by means of
the expansion
\begin{equation}
 \tilde{\Psi}_d(r_{\perp},z,s) = \sum_{\ell,m} R_\ell(r) {\cal Y}_{\ell m}(\theta,\phi) \;,
\label{eq14}
\end{equation}
where the radial function reveals the weight of each partial wave.  

As discussed in section \ref{sec2}, in our case of a squeezing potential
with axial symmetry, $m$ is a good quantum number, and since only
$s$-waves are considered, we then have $m=0$, and the partial wave
components in the wave function (\ref{eq14}) will depend only on 
$r$ and the polar angle $\theta=\arctan(r_\perp/z)$.
It is therefore evident how any observable can be computed via only
$d$-calculated quantities.

Summarizing, given a two-body system interacting via a potential
$V(r)$ with scattering length $a_{3D}$, and put under the effect of
an external potential with oscillator length $b_{ext}$, the procedure
is as follows: i) Calculate the key quantities, i.e. rms radius of the bound 
state in two dimensions, and the 2D scattering length when $a_{3D}=\infty$.
ii) get the corresponding dimension $d$ from the
analytic expression (\ref{enerfit}), or Fig.~\ref{fig2}b, iii) obtain
$\Psi_d$ and $r_d$ in Eqs.(\ref{eq8}) and (\ref{eq9}) after solving
the Schr\"{o}dinger Eq.(\ref{schd}) for that particular dimension, iv)
find the scale parameter, $s$, from the analytic expression in
Eq.(\ref{eq12}), or Fig.~\ref{fig4}b, by use of the ratio of
$b_{ext}/r_d$, and finally, v) use the wave function (\ref{eq14}) in
order to compute whatever observable is of interest.

It is perhaps fair to issue the warning that choosing another external
squeezing field may change the results.  We anticipate that the
scaling parameter is constant, independent of $z$, because the
external oscillator potential is quadratic in $z$, that is our
scaling of the squeezed coordinate. For
another potential a new investigation must be carried out, but such
generalization to other shapes is far beyond the scope of the present
letter.

\section{Perspectives and Conclusions}

Squeezing by an external field and a dimension ($d$) dependent
centrifugal barrier are two formulations to account for a continuous
variation of the spatial dimension between ordinary integer values. We
have demonstrated that the methods are equivalent for two-body systems
with a unique and universal correspondence between external field and
the $d$-parameter.  The relations are relatively simple and can be
expressed approximately by an analytic formula.  The $d$-formulation is
much simpler, both smaller in computer space and time, and therefore
more efficient.  In fact, the method is precisely as complicated as an
ordinary three dimensional calculation with a centrifugal barrier,
that is using only relative coordinates. Thus, the wave function 
with an external field in an ordinary 3D calculation is easily obtained.

Clearly this is not of great consequence for simple two-body systems.
However, the brute force method with squeezing by an external field
is already much more difficult for a three-body system than the
already mathematically available $d$-method.  The interpretation and
equivalence to an external field only has to be worked out and
tested.  However, this is not an easy matter and it requires an
investigation on its own.  Especially mass asymmetric systems with
several and more complicated excited states is a challenge and must be
handled.

We treated the most important transition from three to two spatial
dimensions,  but different continuous variation of dimension can be 
suggested, like from two to one and further on down to zero dimension.
Beside doing the squeezing on one dimension at a time it is possible
to go directly from three to one (or zero) by for example squeezing
two (or all three) directions simultaneously.  Then the integer
dimension of two (or one) would never be passed, but the
$d$-formulation is still valid although the interpretation and the
equivalence translation must change.  The actual calculations are in
all these cases still equally simple and dealing with only relative
coordinates. 

The key to the technical simplifications is the translation from
the external field to the $d$-formulation.  We provide a simple and direct
guide to go from $d$-calculations to ordinary three dimensional
quantities, still only using $d$-quantities like energies and wave functions. 
 We have chosen to
investigate in details the use of an oscillator field as it appears to
be both relatively simple and also practically achievable in
laboratories.  However, other external fields can also be studied as
for example a square well squeezing where the interior of the squeezed
directions have zero potential and only the walls cause the
dimensional confinement.  As the experimental techniques rapidly
develop other shapes may come in as interesting.

In summary, we have for a simple system demonstrated how the efficient
calculations with the $d$-formulation are equivalent to the more
complicated results obtained by squeezing with an external field.  The
energies and wave functions obtained with the $d$-method are for each
$d$ demonstrated to be uniquely related to the parameter, $b_{ext}$,
of the external field.  We illustrated by using both Gaussian and
Morse potentials.  Thus, the $d$-method provides the same full
information as a brute force three dimensional calculation with the
external field.

\section*{Acknowledgements}
This work has been partially supported by the Spanish
Ministerio de Econom\'{\i}a y Competitividad under Projects FIS2014-51971-P,
FPA2015-65035-P and MTM2015-63914-P.


\begin{thebibliography}{0}
\bibitem{gaus1} The one-gaussian approximation is the best variational wave function with only one gaussian minimizing the expectation value of a given hamiltonian.
\end{thebibliography}
\end{document}